\documentclass[onecolumn,showpacs,preprintnumbers,amsmath,amssymb,APSl,nofootinbib,prd]{revtex4}
\usepackage[breaklinks]{hyperref}
\usepackage[english]{babel}
\usepackage{graphicx}
\usepackage{color}
\usepackage{amsfonts,amsthm}
\usepackage{bm,bbm}
\usepackage{mathtools,slashed}

\newcommand{\dm}{\ensuremath{\underline{\gamma}}}

\usepackage{comment}

\newcommand{\mn}{\ensuremath{{_{\mu\nu}}}}

\newcommand{\pa}[1]{\ensuremath{\partial_{#1}}}

\newcommand{\al}{\ensuremath{\alpha}}
\newcommand{\be}{\ensuremath{\beta}}
\newcommand{\ga}{\ensuremath{\gamma}}
\newcommand{\Ga}{\ensuremath{\Gamma}}

\newcommand{\na}{\ensuremath{\nabla}}

\newcommand{\bpsi}{\ensuremath{\bar{\psi}}}

\newcommand{\lnm}{\ensuremath{\Lambda_{Q}}}

\newcommand{\ded}[2]{\ensuremath{{\delta_{#1}}^{#2}}}

\newcommand{\viu}[2]{\ensuremath{{e^{#1}}_{#2}}}
\newcommand{\vid}[2]{\ensuremath{{e_{#1}}^{#2}}}

\newcommand{\lag}{\mathcal{L}}

\newcommand{\lr}[1]{\left(#1\right)}
\newcommand{\lrsq}[1]{\left[#1\right]}

\newcommand{\ee}{\end{equation}}
\newcommand{\ba}{\begin{eqnarray}}
\newcommand{\ea}{\end{eqnarray}}
\def\bs{\begin{subequations}}
\def\es{\end{subequations}}

\def\cO{\mathcal{O}}

\newcommand{\opar}{\overset{\text{\scriptsize$\leftrightarrow$}}{\partial}}
\newcommand{\ospar}{\overset{\text{\scriptsize$\leftrightarrow$}}{\slashed{\partial}}}

\def\com{\color{magenta}}
\def\cob{\color{blue}}
\newcommand{\book}[5]{\emph{#1} (#2, #3, #4, #5)}
\newcommand{\oarX}[1]{\href{http://arxiv.org/abs/#1}{{\ttfamily\com arXiv:#1}}}
\newcommand{\arX}[1]{\href{http://arxiv.org/abs/#1}{{\ttfamily\com arXiv:#1}}}
\newcommand{\doin}[6]{\href{http://dx.doi.org/#1}{{\cob #2 #3 {\bf #4}, #5 (#6)}}}
\newcommand{\doinn}[5]{\href{http://dx.doi.org/#1}{{\cob #2 {\bf #3}, #4 (#5)}}}
\newcommand{\doij}[5]{\href{http://dx.doi.org/#1}{{\cob #2 #3 (#5) #4}}}

\newcommand{\tia}[1]{}

\newcounter{listcounter}

\begin{document}

\title{Observable traces of non-metricity: new constraints on metric-affine gravity }

\author{Adria Delhom }
\email{adria.delhom@uv.es}
\affiliation{Departamento de F\'{i}sica Te\'{o}rica and IFIC, Centro Mixto Universidad de
Valencia - CSIC. Universidad de Valencia, Burjassot-46100, Valencia, Spain}

\author{Gonzalo J. Olmo}
\email{gonzalo.olmo@uv.es}
\affiliation{Departamento de F\'{i}sica Te\'{o}rica and IFIC, Centro Mixto Universidad de
Valencia - CSIC. Universidad de Valencia, Burjassot-46100, Valencia, Spain}

\author{Michele Ronco}
\email{michele.ronco@roma1.infn.it}
\affiliation{Dipartimento di Fisica, Universit\`{a} di Roma ``La Sapienza", P.le A. Moro 2, 00185 Roma, Italy}
\affiliation{INFN, Sez.~Roma1, P.le A. Moro 2, 00185 Roma, Italy}

\begin{abstract}
Relaxing the Riemannian condition to incorporate geometric quantities such as  \textit{torsion} and \textit{non-metricity} may allow to explore new physics associated with defects in a hypothetical space-time microstructure. Here we show that non-metricity produces observable effects in quantum fields in the form of 4-fermion contact interactions, thereby allowing us to constrain the scale of non-metricity to be greater than 1 TeV by using results on Bhabha scattering. Our analysis is carried out in the framework of a wide class of theories of gravity in the metric-affine approach. The bound obtained represents an  improvement of several orders of magnitude to previous experimental constraints.
\end{abstract}

\maketitle

\section{ Introduction}  
The Einstein equivalence principle is the cornerstone of gravitational physics, supporting the idea that gravitation can be interpreted as a geometric phenomenon. Particles and radiation fields follow special paths determined by a curved geometry, whereas their local causal relations are determined by the metric. 
General Relativity (GR) and, more generally, metric theories of gravity are built under the assumption that the geometry is (pseudo-)Riemannian, i.e., that  
{the metric fully determines the spacetime geometry}. However, the experimental limits on the Riemannian assumption are still not well established at short length (high energy) scales \cite{will,berti,daniel,henric,thib,kramer,barrow,gw} and  some steps should be taken to better understand whether geometric structures other than the metric could be needed to account for all space-time properties.  In this regard, it is worth noticing that though much has been done to infer the potential existence of higher dimensions \cite{servant} or supersymmetry \cite{atlas}, the roles of torsion \cite{tor} and, specifically, of non-metricity \cite{nonm} {in the high-energy (short-distance) regime are less known}. As a reminder, torsion is the antisymmetric part of the connection and non-metricity is a tensor that measures the failure of the connection to be compatible with the metric. We use the definitions in \cite{Ortin}: ${S_{\mu\nu}}^\lambda\equiv-2{\Gamma_{[\mu\nu]}}^\lambda$ is the torsion tensor, and $Q_{\lambda\mu\nu}\equiv-\nabla_\lambda g\mn$ is the non-metricity tensor. {The physical relevance of these magnitudes can be appreciated in condensed matter systems with a microstructure. In fact, in the continuum limit, the lattice structure of a solid gives rise to an effective (or emergent)  continuous geometry which cannot be described solely in terms of an effective metric \cite{cond1,Clayton,cond2}. Torsion and non-metricity become necessary to fully account for the physical characteristics of those systems, such as plasticity and viscoelasticity, which are intimately related with the presence of topological defects in the crystal lattice \cite{Clayton,cond1,cond2,elast,geon1}}. 
Given our limited understanding of gravitation in the high-energy regime, if the continuous space-time that we perceive had some kind of  ``microstructure", as expected in almost the totality of approaches to quantum gravity \cite{Ori09,Fousp,Zwi09,rov07,thi01,Per13,gacLRR,GiSi,AGJL4,Dow13,LaR5,NiR,RSnax,ADKLW,BIMM,Hor3,CES,Tom97,Mod1,BGKM,CaMo2,frafuzz1,frafuzz2,sabine}, it is legitimate to explore how the continuum may arise and the potential impact that geometrical objects other than the metric could have in a low-energy effective description. Though some observable effects of torsion have been experimentally tested \cite{mao,simone,peron,lorenzo,lehn}, the observable consequences of non-metricity  still represent a largely unknown territory.

The main purposes of this work are 1) to show that the observable effects of non-metricity may be more easily accessible than those of torsion and 2) set a lower limit to the energy scale $\Lambda_{Q}$ at which non-metricity may become important. To this end, we consider the effects of non-metricity in quantum systems involving spinor fields in both relativistic and non-relativistic scenarios. To carry out this study, it is first necessary to work out the generalization of the covariant Dirac equation \cite{BirrellDavies} for space-times with the most general form of non-metricity. The nontrivial new elements involved in this generalized equation of motion already suggest that non-metricity could have observable effects, but a precise quantitative analysis requires the definition of specific forms of non-metricity. For this purpose, we use as a guide the predictions of a wide class of metric-affine theories of gravity recently studied in the literature \cite{Afonso:2017bxr}, which we call RBG (Ricci-Based Gravity) theories. These theories are defined by actions of the form
\begin{equation}\label{eq:Action0}
S(g\mn,{\Gamma_{\mu\nu}}^\alpha,\psi_m)=\int d^4 x\sqrt{-g} f_{RBG}(g\mn,R_{(\mu\nu)}) +S_m \ ,
\end{equation}
 where $R_{(\mu\nu)}$ denotes the symmetrized Ricci tensor, $g\mn$ the space-time metric, and $\psi_m$ the matter fields that appear in $S_m$. Here $R\mn\equiv{R^\alpha}_{\mu\al\nu}$, where the Riemann tensor is ${R^\alpha}_{\beta \mu\nu}=\partial_\mu{\Gamma_{\nu\beta}}^\alpha -\partial_\nu{\Gamma_{\mu\beta}}^\alpha+{\Gamma_{\mu\lambda}}^\alpha{\Gamma_{\nu\beta}}^\lambda-{\Gamma_{\nu\lambda}}^\alpha{\Gamma_{\mu\beta}}^\lambda$, and ${\Gamma_{\mu\beta}}^\alpha$ is the connection (a priori independent of the metric $g_{\mu\nu}$). The Lagrangian $f_{RBG}(g\mn,R\mn)$ is actually built in terms of traces of powers of the object $g^{\mu\alpha}R_{\alpha\nu}$ in order to guarantee that it is a scalar function and that GR is recovered as a low-energy limit in a power series expansion \cite{Jimenez:2014fla}. Corrections to GR appear as terms proportional to inverse powers of $\Lambda_{NM}$, the scale which we here wish to constrain experimentally. Our results apply to the vast majority of metric-affine theories studied so far, e.g.  $f(R)$, Born-Infeld (BI), and other extensions (see  \cite{olmo1,beltran,cap,tsu,sotiriou} for related reviews). Indeed, a common feature of all RBG models (with minimal matter couplings) is that non-metricity is sourced by the local densities of energy and momentum  \cite{beltran,breton,paolo,komada,hendi,janakar,avelino,rinaldi,beltran1,psisnchen,makOdinOlmo,clifton,vollick1,vollick2,DMbi,brw}. We explicitly show that this leads to geometry-induced interactions among the matter fields. As a result, to first order in perturbation theory, this coupling generates $4$-particle fermion interactions whose amplitude can be orders of magnitude larger than the $4$-fermion interactions typically associated with torsion. Within this approximation, the effects of torsion in RBGs are the same as in GR and, thus, can be consistently neglected. Given that, by using experimental data for Bhabha scattering from LEP \cite{ExpData1,ExpData2}, we are able to constrain the energy scale at which non-metricity effects may arise, setting a lower bound for $\lnm$ at the TeV scale.

Gravity-induced point-like interactions were already pointed out by Flanagan \cite{flanagan} (see also \cite{VollickFlan}) in the context of $1/R$ gravity using a scalar-tensor representation. The existence of a certain gauge freedom associated to the projective invariance of RBGs, however, does not allow to interpret those effects in $f(R)$ theories as due to the \textit{non-metricity}, as a metric-compatible gauge choice (with torsion) of those theories is possible \cite{Afonso:2017bxr,Capozziello:2007tj,Sotiriou:2009xt}. On the contrary, the  broader class of gravity theories considered here possesses genuine \textit{non-metricity}, and we are able to ascribe the appearance of additional particle contact interactions directly to this term [see below Eq.(\ref{nonmetricitystressenergy}) for more details]. 
 
 The effects of non-metricity have  also been considered recently in the context of Lorentz symmetry breaking assuming that the non-metricity tensor has a constant non-zero vacuum expectation value \cite{kostelecky}. In the framework of RBGs, as we will see, this condition requires a nonzero vacuum expectation value $\nabla_\alpha \langle T_{\mu\nu}\rangle$, where $T_{\mu\nu}$ is the stress-energy tensor of the matter fields. This is so because in these theories the non-metricity turns out to be an induced quantity fully determined by the matter rather than an independent  dynamical field. The conditions under which the results of   \cite{kostelecky} hold, therefore, seem to be more restricted than originally expected.

\section{Field equations}  
  
Let us begin by discussing the field equations that follow from the action (\ref{eq:Action0}). As shown in \cite{Jimenez:2014fla,Afonso:2017bxr,Emanuele}, Einstein field equations get modified in this family of theories as follows
\begin{equation}\label{EinsteinlikeEqs}
{G^\mu}_\nu(\Gamma)=\frac{\kappa^2}{|\Omega|^{1/2}}\left[{T^\mu}_\nu-\delta^\mu_\nu\left(f_{RBG}+\frac{T}{2}\right)\right] \ ,
\end{equation}
where ${T^\mu}_\nu\equiv g^{\mu\alpha}T_{\alpha\nu}$, $T={T^\mu}_\mu$, ${G^\mu}_\nu(\Gamma)\equiv q^{\mu\alpha}G_{\alpha\nu}(\Gamma)$, with $q_{\mu\nu}$ representing an auxiliary metric defined by the relation $\sqrt{-q}q^{\mu\nu}\equiv \frac{1}{2\kappa^2}\sqrt{-g}{\partial f_{RBG}}/{\partial R_{\mu\nu}}$, and $|\Omega|^{1/2}\equiv \sqrt{-q}/\sqrt{-g}$. If the matter fields are not coupled to the independent connection, one finds that torsion is trivial and $\Gamma^\alpha_{\mu\nu}$ is the Levi-Civita connection associated with $q_{\mu\nu}$.  When matter fields do couple to the connection torsion terms may arise, though they turn out to play a negligible role in our discussion, as will be explained later.
 One can then find an on-shell relation of the form $ g_{\mu\nu}=q_{\mu\al}{(\Omega^{-1})^\al}_\nu$, where ${(\Omega^{-1})^\al}_\nu$ can be expanded as a series in powers of the stress-energy tensor of the form\footnote{On-shell, when $\Ga=\Ga(q\mn)$, one can also write $f_{RBG}$ as a function of the matter fields and the auxiliar metric \cite{Cracking,beltran}.} \cite{Cracking,beltran}
\begin{equation}\label{OmegaSeries}
{(\Omega^{-1})^\al}_\nu={\delta^\al}_\nu+\frac{1}{\lnm^4}\lr{\al T{\delta^\al}_\nu+\be{T^\al}_\nu}+\mathcal{O}(\lnm^{-8})
\end{equation}
where $\alpha$ and $\be$ are numerical coefficients whose specific value depends on the particular RBG model under consideration (usually of order $\mathcal{O}(1)$). In RBG models \cite{beltran}, this energy scale takes the form\footnote{Here $E_p$ is the Planck energy and $\Lambda_{RBG}=hc/\lambda_{RBG}$.} $\lnm=(8\pi G \lambda_{RBG}^2)^{-1/4}=(2\pi)^{1/4}(E_p\Lambda_{RBG})^{1/2}$ in units $\hbar=c=1$; where $\Lambda_{RBG}$ is a high-energy scale that characterizes RBG models (the difference between the models is given by different values of $\al$ and $\be$). Rather than finding these model-dependent coefficients, we here aim to constrain the general energy scale $\lnm$. Finally, let us notice that from \eqref{OmegaSeries} it follows an equivalence between all RBG and GR in the vacuum, where $T\mn=0$.

In order to explore the physics associated to the break-down of the metricity condition, $\nabla_\alpha g_{\mu\nu}=0$, a non-metricity tensor must be specified. In this sense, RBG theories are a rather general but  sufficiently simple family that nicely fits to our purposes  \cite{Afonso,hehl1,hehl2,plbTor,Palageod,PalaNNE,Avelino:2012qe,neutronstars}. Whithin these theories, using \eqref{OmegaSeries} one can write the metric and non-metricity tensors to lowest order in $1/\lnm$ in the form\footnote{Notice that $q_{\mu\al}{T^\al}_\nu=g_{\mu\al}{T^\al}_\nu+\cO(\lnm^{-4})$. Thus, up to lowest order in $1/\lnm$, there is no ambiguity in the $T\mn$ of eqs. \eqref{metricstressenergy} and \eqref{nonmetricitystressenergy}, since the term with $T\mn$ is already of order $\lnm^{-4}$ in both equations.}

\begin{eqnarray}
g_{\mu\nu} &=& q_{\mu\nu} +\frac{1}{\lnm^4}\lr{ \al T q_{\mu\nu}+\beta T\mn} \label{metricstressenergy}\\
Q_{\al\mu\nu} &=&- \frac{1}{\lnm^4}\lr{\al(\na_{\al}T) q\mn+\beta \na_{\al}T\mn} \ . \label{nonmetricitystressenergy} 
\end{eqnarray}
 Though the $\alpha$-dependent term in (\ref{nonmetricitystressenergy}) can be gauged away by a projective transformation, as in $f(R)$ theories \cite{Afonso:2017bxr}, the $\beta$ contribution is a genuine form of non-metricity. The effects of $\alpha$, nonetheless, may still arise through the tetrads associated to (\ref{metricstressenergy}).

 The weak field limit of equation \eqref{EinsteinlikeEqs} gives $q_{\mu\nu}\approx \eta_{\mu\nu}+\delta q_{\mu\nu}$, where $\delta q_{\mu\nu}$ accounts for the usual Newtonian and post-Newtonian corrections. This can be seen by substituting $\Ga=\Ga(q\mn)$ in \eqref{EinsteinlikeEqs} and realizing that these are a set of Einstein-like differential equations for the auxiliary metric $q\mn$ as a function of the matter fields. Thus $q\mn$ accounts for effects of integration over the sources (i.e. Newtonian and post-Newtonian; see  \cite{Olmo:2006zu} for an illustration). As we will study non-metric effects in experiments in which Newtonian and post-Newtonian corrections are neglected, hence $q_{\mu\nu}\approx \eta_{\mu\nu}$, our scenario consists of a Minkowskian background corrected by terms that give non-metricity.  Notice that, as non-metricity is sourced locally by (derivatives of) the stress-energy tensor, then how energy and momenta are distributed locally plays a pivotal role in determining short-distance physics, since large departures from the flat Minkowski metric could arise at high enough densities, where (quantum) field fluctuations may be important. This opens the way to the search for departures from GR in the {\it high-energy-density} regime rather than in the high-energy regime, as also suggested elsewhere relying on different arguments \cite{planckstar}. It has been suggested that this could have observable consequences in different scenarios with high energy density, such as stellar and nuclear matter models \cite{beltran}.

\section{Physical effects of non-metricity}

 In this section, we show that non-metric effects may be important in the context of particle collisions and atomic physics by studying the corrections that non-metricity introduces in fermion-fermion scattering and in atomic spectra. This allows us to propose tests to non-metricity in both relativistic and non-relativistic regimes. The results will be used to estimate  the impact that an underlying space-time structure with non-metricity described by an RBG might have in particle physics. This will allow us to constrain the energy scale\footnote{The constraint will generally depend on $\alpha$ and $\beta$, which are model dependent but usually of order $\cO(1)$.} $\Lambda_{RBG}$ at which RBG models are compatible with experiments, and also the scale $\lnm$ above which a non-metricity like \eqref{nonmetricitystressenergy} could still describe the space-time geometry according to current experimental data. 

\subsection{Effective 4-fermion interactions}  

Let us show how non-metricity within RBG induces effective interactions between a pair of spin 1/2 fields and any other fields existing in nature. For this purpose, let us start with the standard covariant Lagrangian for a spin 1/2 field in a curved spacetime \cite{BirrellDavies}
\begin{equation}\label{BDlag}
\lag=\sqrt{-g}\lrsq{\frac{1}{2}\lr{\bpsi\dm^\mu(\na_\mu\psi)-(\na_\mu\bpsi)\dm^\mu\psi}+\bpsi m \psi},
\end{equation}
where $\na_\mu\psi\equiv\lr{\pa{\mu}-\Ga_\mu}\psi$, where $\Ga_\mu\equiv {\omega_{\mu}}^{ab}\sigma_{ab}$ is the spinor connection, ${\omega_{\mu}}^{ab}\equiv \frac{1}{2}\lr{\pa{\mu}\viu{b}{\al}+\viu{b}{\be}{\Ga_{\mu\al}}^\be}\eta^{ac}\vid{c}{\al}$, $\sigma_{ab}\equiv-\frac{1}{4}\lrsq{\ga_b,\ga_a}$\footnote{$\sigma_{ab}$ are the generators of the Lorentz group in its usual form within the spin representation.}and  $\dm^\mu=\vid{a}{\mu}\ga^a$, being $\vid{a}{\mu}$ the vierbein defined by $\vid{a}{\mu}\vid{b}{\nu}g\mn=\eta_{ab}$ and $\ga^a$ the standard Dirac matrices.

Using now the form of the metric \eqref{metricstressenergy} we find to lowest order in $1/\lnm$
\begin{equation}\label{vieconn}
\begin{split}
&\viu{a}{\mu}={\delta^a}_\mu+\frac{\al}{2\lnm^4}T{\delta^{a}}_\mu+\frac{\be}{2\lnm^4}{T^a}_{\mu}\qquad\qquad\vid{a}{\mu}={\delta_a}^\mu-\frac{\al}{2\lnm^4}T{\delta_{a}}^\mu-\frac{\be}{2\lnm^4}{T_a}^{\mu}\\
&\sqrt{-g}=1+\frac{4\al+\be}{2\lnm^4}T\hspace{2,6cm}\Ga_{\mu}=\mathcal{O}({S_{\mu\nu}}^\lambda) \ .
\end{split}
\end{equation}
Now let us make the following consideration: for scattering experiments at the surface of the Earth (concretely in LEP), we can neglect Newtonian and post-Newtonian corrections to the Standard Model Lagrangian. This is equivalent to use $q\mn\approx\eta\mn$ as explained above. By use of \eqref{vieconn} we can thus write the Lagrangian \eqref{BDlag} as $\lag=\lag^0+\lag^I$, where $\lag{}^0$ is the usual Lagrangian for spin 1/2 fields in Minkowski spacetime \cite{BirrellDavies} and $\lag^I$ can be seen as an interaction Lagrangian for a Dirac field in Minkowski spacetime with the stress energy tensor, which takes the form\footnote{$\bpsi\opar_\mu\psi\equiv\frac{1}{2}\lrsq{\bpsi(\pa{\mu}\psi)-(\pa{\mu}\bpsi)\psi}$}
\begin{align}\label{IntLagT}
\lag^I=\frac{\be}{2\lnm^4}\lr{T\lrsq{\bpsi\ospar\psi+\bpsi m\psi}+{T_a}^\mu\lrsq{\bpsi\ga^a\opar_\mu\psi}}+\frac{3\al}{2\lnm^4}T\lrsq{\bpsi\ospar\psi}+\mathcal{O}(\lnm^{-8}).
\end{align}
Here torsion has been neglected because, as shown in \cite{Hehl}, torsion-induced interactions are beyond experimental reach unless a very-high density of spin (the source of torsion \cite{Kibble}) is considered. This behavior of torsion contrasts with that of non-metricity, since the latter is sourced by the energy-momentum density, which can be more easily controlled and magnified in particle colliders. 

The Lagrangian $\lag^I$ evidences that non-metricity in RBGs induces contact interactions between a fermion pair and any kind of field entering the stress-energy tensor (even self-interactions). Accordingly, we can constrain $\lnm$ by requiring that the non-metric contribution to the cross-section of particle processes does not exceed the measurement error. This also implies that theories with non-metricity of the form \eqref{nonmetricitystressenergy} should be regarded as effective theories because the lack of new dynamical degrees of freedom (as compared to GR) together with the existence of 4-fermion contact interactions \eqref{IntLagT} may lead to unitarity violations at the scale $\lnm$ (unless some strong coupling mechanism beyond the linear approximation fixes this issue\footnote{In some RBGs black hole and cosmic singularities may be avoided in a non-perturbative way\cite{nonsingBH1,nonsingBH2}.}). \\
Let us now focus on the process  $e^+e^-\rightarrow e^+e^-$ in the ultra-relativistic regime ($m_e\approx 0$) for which up to lowest order in  $1/\lnm^4$, the Lagrangian \eqref{IntLagT} reads
\begin{equation}\label{Lag2ferm}
\begin{split}
\mathcal{L}^I=-\frac{\be}{\lnm^4}\lrsq{\bpsi_e \lr{\ga_a\overset{\text{\scriptsize$\leftrightarrow$}}{\partial^\mu}+\ga^\mu\opar_a}\psi_e}\lrsq{\bpsi_e\ga^a\opar_\mu\psi_e}.
\end{split}
\end{equation}
Within the Standard Model, the contribution of \eqref{Lag2ferm} to the  cross section of this process at tree level and lowest order in $1/\lnm^4$is
\begin{equation}\label{crossection}
\sigma_{Q} \simeq \,  0.35\frac{\beta}{\lnm^4}  \;\text{pb.}
\end{equation}
Current data on the process $e^+e^-\rightarrow e^+e^-$ can be found in \cite{ExpData1,ExpData2}. Measurements from LEP\footnote{Let us mention that using LHC data for process of the type $q\overline{q}\rightarrow f\overline{f}$ would not improve the limit we here establish. See e.g. \cite{ExpData3}.} at a center of mass energy of $\sqrt{s}=207$ GeV show that the cross section for this process is $\sigma_{exp}=256.9\pm1.4\pm1.3$ pb\footnote{We use the data with $\theta_{acol}<10^o$ and $|\cos\theta_{e^\pm}|<0.96$  \cite{ExpData1,ExpData2}.}  \cite{ExpData1,ExpData2}.  The requirement that any RBG model in the metric-affine approach has to be consistent with current data\footnote{This is done by requiring $\sigma_{SM}+\sigma_{NM}$ is compatible with the experimental value.} sets a lower (upper) bound for $\lnm$ ($\lambda_{Q}$) of about
\begin{align}
&\lnm \gtrsim 0.6 \, {\beta^{1/4}} \,  \text{TeV},\\
&\lambda_{Q} \lesssim 2 \,  {\beta}^{-1/4}\times{10^{-18}} \; \text{m} \, , 
\end{align}
where $\beta$ is the model dependent coefficient appearing on \eqref{OmegaSeries} usually of order $\mathcal{O}(1)$. Correspondingly, for $\Lambda_{RBG}$ ($\lambda_{RBG}$) we have
\begin{align}\label{NMlimit}
&\Lambda_{RBG}\gtrsim 0.06\, \beta_{model}^{1/2} \,  \text{meV},\\
&\lambda_{RBG}\lesssim 2 \, \beta_{model}^{-1/2} \; \text{cm}.
\end{align}
In particular, for BI inspired models with Lagrangian $\left(|\det(\delta^\mu_\nu+\lambda_{BI}^{2} g^{\mu\alpha}R_{\alpha\nu})|^n-1\right)/(8\pi G \lambda_{BI}^2)$ \cite{Odintsov:2014yaa}, one has $\beta_{BI}=\frac{1}{2n}$, with $n=1/2$ corresponding to the so-called Eddington-inspired BI model. Picking out $n=1/2$, the above bounds translate into $\Lambda_{BI} \gtrsim 0.06 \, $ meV and $\lambda_{BI} \lesssim 2\,$ cm. It is worth mentioning that these bounds we here established are in the range recently highlighted in the naive estimations of    \cite{beltran}.  Let us stress that this represents an improvement on the previous best limit on $\lambda_{BI}$ (see e.g. \cite{Avelino:2012qe,neutronstars}) by more than 7 orders of magnitude. A worth feature of the above constraint is that it weakly depends on the details of the model considered. For astrophysical and cosmological bounds on the $n=1/2$ model see  \cite{beltran}.

\subsection{Shifts in the Energy levels of atomic systems}

Let us now illustrate a preliminary proposal to perform tests also in the non-relativistic regime. As first shown by Parker \cite{parker1,parker2,parker3}, strong gravitational fields produce modifications in the atomic interaction Hamiltonian, which then induces specific shifts in their energy levels in regions of high curvature. One may wonder if similar effects could be sourced by high-density concentrations through non-metricity.

 Parker starts with the usual Dirac equation in curved spacetimes, which is known as the Dirac-Weyl-Fock (DWF) equation, and it reads \cite{parkerBook}
\begin{align}\label{direq}
\lr{\dm^\mu\nabla_\mu+m}\psi=0.
\end{align}
 In non-Riemannian spacetimes, the above equation \eqref{direq} gets corrected by terms involving non-metricity and torsion. This is due to two reasons: 1) that the relation between the divergence operator and the covariant derivative is not the usual one  \cite{PaperDivergence}, and 2) that the curved Dirac matrices are no longer covariantly constant. This can be seen in more detail in \cite{PaperDivergence,Formiga}, where the DWF equation in spaces with general non-metricity and torsion is found to be
\begin{align}\label{DirModEq}
\lrsq{\dm^\mu\na_\mu+\frac{1}{2}\lr{{S_{\mu\al}}^\al+{Q_{[\al\mu]}}^\al}\dm^\mu+m}\psi=0,
\end{align}
Notice that for Riemannian spaces \eqref{direq} is recovered as expected. Following Parker's approach, a non-metric interaction Hamiltonian can be defined from \eqref{DirModEq} by identifying $H_D=i\pa{t}$ and stating $H_I\equiv H_D-H_D^M$, where $H_D^M$ is the Dirac Hamiltonian in Minkowski spacetime. This leads to the following interaction Hamiltonian for fermions in a curved and non-metric spacetime:
\begin{eqnarray}\label{InteractionHamiltonian}
H_I=-i\left[\ga^a\ga^b\lr{\frac{\vid{a}{0}\vid{b}{i}}{g^{00}}+\delta_a^0\delta_b^i}\pa{i}+\frac{\vid{a}{0}\vid{b}{i}}{g^{00}}\lr{q_i-\Ga_i}\right.+\left.q_0-\Ga_0+\lr{\frac{\vid{a}{0}}{g^{00}}+\delta_a^0}m\ga^a\right] \ , 
\end{eqnarray}
where we defined $q_{\mu} \equiv 1/2\lr{{S_{\mu\al}}^\al+{Q_{[\al\mu]}}^{\alpha}}$. Using \eqref{vieconn} and neglecting again torsion for the aforementioned reasons, up to lowest order in $1/\lnm^4$ we find
\begin{equation}\label{IntHamTmunu}
H_I=\frac{i}{\lnm^4}\left[-\beta\lr{T^{00}\ded{a}{0}\ded{b}{i}+\frac{1}{2}\lr{\ded{a}{0}{T_b}^i+{T_a}^0\ded{b}{i}}}\ga^a\ga^b\pa{i}+\right.\frac{3\al+\be}{4}(\pa{a}T)\ga^0\ga^a+\left.\lr{\frac{\al}{2}T\ded{a}{0}-\frac{\be}{2}{T_a}^0-\be{T^{00}}\ded{a}{0}}m\ga^a\right].
\end{equation} 
Keeping only the leading order terms in the non-relativistic limit as in \cite{parker3}\footnote{$\ga^i=-i\tilde{\beta}\tilde{\alpha^i}\sim\mathcal{O}(\alpha_{em}),\;\ga^0=-i\tilde{\beta}\sim\mathcal{O}(1),\;\pa{i}\sim p_i\sim\mathcal{O}(\alpha_{em})$}, one finally has
\begin{equation}\label{IntHamTmunuNR}
H_I^{NR}=-i\frac{3\al+\be}{4\lnm^4}\pa{0}T-\frac{m}{2\lnm^4}\lr{\be T^{00}-\al T}.
\end{equation}
In order to test non-metricity effects through energy shifts of atomic levels, one should be able to change the local distributions of energy and momentum around the atom minimizing the impact of undesired electromagnetic couplings. Clouds of dark matter particles and/or intense neutrino fluxes, both having very weak or no couplings to the electromagnetic sector, could do the job. 
We will work out the case of a Hydrogen atom traversed by a radiation flux emitted by a spherically symmetric source. This fluid is modeled as 
an ideal null fluid with\footnote{Here $\rho$ is the energy density of the fluid and $l_\mu$ is a radial null vector.} $T\mn=\rho l_\mu l_\nu$. Plugging this into  \eqref{IntHamTmunuNR}, the non-relativistic interaction Hamiltonian turns into 
\begin{equation}\label{EffectiveNRHamiltonian}
H_{I}^{NR}=-\frac{\beta}{2\lnm^4}m\rho \ .
\end{equation}
Assuming an energy density profile that decays with the distance $R$ to the center of the source as $\rho = 4\rho_s R^2_s/(R_s + R)^2$, being $R_s$ the size of the source and $\rho_s$ the energy density of the flux at $R=R_s$, the non-metricity correction to the energy levels is 
\begin{equation}\label{shift1}
\Delta^Q_{(n,l,m)} \simeq -\frac{\beta}{2\lnm^4}m\rho_s\left( 1+ \frac{1}{R^2_s}\left< 4r^2 \cos^2\theta -r^2 \right>_{nlm} \right) ,
\end{equation}
where $r$ measures the distance from the center of the atom, and terms of order $(r/R_s)^3$ and higher have been neglected. Then, for a state of the form  $(n,0,0)$, one  gets
\begin{equation}\label{shift2}
\Delta^Q_{(n,0,0)} = -\frac{\beta}{2\lnm^4}m\rho_s\left( 1- \frac{1}{3}\left(\frac{na_0}{R_s}\right)^2(5n^2 +1)\right) \, , 
\end{equation}
being $a_0$ the Bohr radius and $m$ the electron mass.  We noticed that the non-metric shift represents an extremely small correction to the unperturbed energy levels. Given that, let us consider an optimistic situation (only for illustrative purposes) by taking high values for $n$ as well as for $\rho_s$ in order to amplify the non-metric correction as much as possible. Consider a very large $n$  transition $(1000,0,0)\to (2,1,0)$ and take a very high energy density $\rho_s \simeq 10^{31} \,$ J/$\text{m}^3$ (comparable in magnitude to the observed gravitational wave events generated by black hole mergers), the constraint that one finds by requiring $\Delta^Q$ to be less than the error $\sim \, \alpha_{em}^2$ due to neglecting relativistic corrections is just $\lambda_{RBG} \lesssim 10^{17}\beta_{model}^{-1/2}$ m. Though this bound is orders of magnitude weaker than our relativistic estimates, it is still a better constraint than the one obtained in \cite{BoundGRWaves} from GW170817 and GRB 170817A. Other avenues to explore non-metricity effects on atomic systems may involve the study of rapid transients, which are sensitive to the coefficient $\alpha$ in (\ref{IntHamTmunuNR}). This requires a more detailed modeling of the sources and the use of time-dependent perturbation theory, aspects which will be explored elsewhere. \\

\section{Conclusions}


 We have shown that for RBG models in the metric-affine approach, non-metricity gives rise to potentially observable effects in microscopic systems, which can be used to impose tight constraints on the model parameters. Though our analysis focuses on the RBG family of theories, one can envision other extensions in which the non-metricity could become a dynamical entity.  If terms such as powers of the Riemann tensor are included, a decomposition in terms of its irreducible components indicates that quadratic Ricci contributions would arise, leading to effects similar to those described here. Our results, therefore, are likely to be a rather generic property of gravity theories with non-metricity.  We feel that recognizing this already provides some additional insight on our understanding of the basic properties of the space-time geometric structure and the potential impact of non-metricity on experiments at short scales. 
  
  Using current data for $\bold{e}^+\bold{e}^-$ scattering, we have been able to set a lower bound of the order of $1$ TeV on the scale at which non-metricity could be present without being in conflict with experiments. We also found that a general consequence of non-metricity in the context of RBGs is the induction of contact 4-particle interactions among a fermion-antifermion pair and any other pair of particle-antiparticle in the Standard Model. Consequently, Higgs physics at LHC or its impact on flavor physics could provide complementary bounds for non-metric effects. Moreover, forthcoming accelerator experiments such as CLIC \cite{clic} will be used to perform high-precision measurements of $\bold{e}^+\bold{e}^-$ collisions with a center of mass energy around the TeV scale (almost one order of magnitude more than the LEP configuration we used). If no departures from the Standard Model are found in those experiments, our bounds might be improved by a few orders of magnitude. 
 
 Also interesting would be understanding the role of non-metricity in the production of non-linearities of the cosmological perturbations that  reflect into non-gaussianities in the Cosmic Microwave Background \cite{planck1,planck2,ivan}. Here we did not explore this possibility, which though should be investigated elsewhere.  At the same time, we hope that non-metricity might open a new window to detect the presence of energy-momentum fluxes carried by weakly interacting sources. In fact, we have here shown that, in RBG models, these fluxes would change the energy levels of atoms in a way that depends on their energy density. The non-relativistic interaction Hamiltonian (\ref{IntHamTmunuNR}) allowed us to explore the impact of such fluxes using atomic energy levels, which unfortunately turned out to be extremely tiny. Nonetheless, we hope that transient processes (able to excite the time derivatives appearing in Eq.(\ref{IntHamTmunu})) and nuclear physics scenarios (with high energy density) are likely to provide more insightful experimental constraints. The rapid progress experienced (and expected) in atomic interferometry could also help explore this sector of gravitational physics through a variety of new experiments.  \\

\paragraph*{Acknowledgments.} 
G.J.O. and A.D.L. are supported by  a Ramon y Cajal contract (RYC-2013-13019) and an FPU fellowship, respectively.
M. R. thanks the Department of Theoretical Physics \& IFIC at the University of Valencia for hospitality and partial support during the elaboration of this work. This work is supported by
the Spanish grant FIS2014-57387-C3-1-P (MINECO/FEDER, EU), the project 
H2020-MSCA-RISE-2017 Grant FunFiCO-777740,
 the Consolider Program CPANPHY-1205388, and the Severo Ochoa grant SEV-2014-0398 (Spain). This article is based upon work from COST Action CA15117, supported by COST (European Cooperation in Science and Technology).We also thank Jos\'e Beltran Jimenez, Joan Ruiz Vidal, and Ivan Rosario Bonastre for fruitful discussions during the elaboration of this work.

\end{document}